\documentclass[fleqn,usenatbib]{mnras}

\usepackage{newtxtext,newtxmath}

\usepackage[dvipsnames]{xcolor}

\usepackage[T1]{fontenc}
\usepackage{ae,aecompl}

\usepackage{graphicx}	
\usepackage{amsmath}	

\title[MRI saturation and magnetically elevated discs]{Saturation of the magnetorotational instability and the origin of magnetically elevated accretion discs}

\author[M. C. Begelman and P. J. Armitage]{
Mitchell C. Begelman$^{1,2}$\thanks{E-mail: mitch@jila.colorado.edu}
and Philip J. Armitage$^{3,4}$
\\
$^{1}$JILA, University of Colorado and National Institute of Standards and Technology, 440 UCB, Boulder, CO 80309-0440, USA \\
$^{2}$Department of Astrophysical and Planetary Sciences, 391 UCB, Boulder, CO 80309-0391, USA \\
$^{3}${Center for Computational Astrophysics, Flatiron Institute, 162 Fifth Avenue, New York, NY 10010, USA} \\
$^{4}${Department of Physics and Astronomy, Stony Brook University, Stony Brook, NY 11794, USA}
}

\date{Accepted 21 March 2023. Received 18 February 2023; in original form 15 November 2022}

\pubyear{2022}

\begin{document}
\label{firstpage}
\pagerange{\pageref{firstpage}--\pageref{lastpage}}
\maketitle

\begin{abstract}
We propose that the strength of angular momentum transport in accretion discs threaded by net vertical magnetic field is determined by a self-regulation mechanism: the magnetorotational instability (MRI) grows until its own turbulent resistivity damps the fastest growing mode on the scale of the disc thickness. Given weak assumptions as to the structure of MRI-derived turbulence, supported by prior simulation evidence, the proposed mechanism reproduces the known scaling of the viscous $\alpha$-parameter, $\alpha \propto \beta_z^{-1/2}$. Here, $\beta_z = 8\pi p_g/B_{z0}^2$ is the initial plasma $\beta$-parameter on the disc midplane, $B_{z0}$ is the net field, and $p_g $ is the midplane gas pressure. We generalize the argument to discs with strong suprathermal toroidal magnetic fields, where the MRI growth rate is modified from the weak-field limit. Additional sources of turbulence are required if such discs are to become magnetically elevated, with the increased scale heights near the midplane that are seen in simulations. We speculate that tearing modes, associated with current sheets broadened by the effective resistivity, are a possible source of enhanced turbulence in elevated discs.
\end{abstract}
 



\section{Introduction}
The magnetorotational instability \citep[MRI;][]{balbus1991,balbus98} is a widely-applicable mechanism that generates moderately strong levels of turbulence and angular momentum transport in accretion discs. Despite detailed linear analyses \citep{pessah05,latter15,das18}, and simulations of increasing numerical fidelity and physical realism \citep[e.g.][]{ryan17,scepi18,jiang20}, some foundational questions remain unresolved. Notably, the MRI saturates in the limit of very weak net fields at a level where the turbulence is highly subsonic, and the magnetic energy density is highly subthermal. For stronger --- but still weak --- fields, the viscous $\alpha$-parameter \citep{shakura73} scales with the net vertical field as
\begin{equation}
    \alpha \propto \beta_z^{-1/2}.
\end{equation}
Here, the mid-plane plasma $\beta$-parameter is defined in terms of the net field $B_{z0}$ and mid-plane gas pressure through $\beta_z = 8\pi p_g/B_{z0}^2$. This relationship is seen in both local  \citep[shearing box:][]{hawley95,bai13,salvesen16} and global \citep{mishra20} MRI simulations, extending over roughly $10 \lesssim \beta_z \lesssim 10^5$, with a typical coefficient $\sim O(10)$. The origin of this scaling is unclear.

The puzzle of this relationship is twofold. First, the dependence on $p_g$ implies that the MRI is aware of the sound speed, despite the fact that for weak fields saturation occurs when $\alpha \sim 10^{-2}$ \citep{davis10,simon12} and the fluid motions are effectively incompressible. Second, the dependence on $\beta_z$ suggests that the level of turbulence is set by MRI operating on sufficiently large scales (e.g., comparable to the disc scale height, $H$) that the mean field cannot be swamped by turbulent fluctuations, despite the fact that shorter-wavelength modes have much shorter growth times when $\beta_z$ is large. The solution to these longstanding conceptual issues may be related to, and inform, more recent problems in our understanding of magnetically elevated discs. These are discs where moderately strong vertical fields ($\beta_z \sim 10^2-10^3$) catalyze the development of stronger, in some cases suprathermal, toroidal fields \citep{bai13,salvesen16,zhu18,lancova19,mishra20}. Magnetically elevated discs are thicker than predicted based on the vertical support from gas or radiation pressure gradients, but the mechanism of this additional support is not well-understood.  While discs with a net vertical magnetic flux develop coherent toroidal and radial fields \citep{begelman07,begelman15,salvesen16}, these fields reverse across the midplane and increase with height roughly linearly across about a scale height on either side.  This means that, within the inner, magnetically supported scale height, the pressure force from these large-scale fields cannot supply the vertical support needed. It is therefore up to the turbulent field to provide such support.
  
In this paper, we propose a self-regulation mechanism for MRI-maintained turbulence that could resolve these conceptual issues, while providing a simple way to estimate the saturated stress for the net-field MRI.  Our idea is based on the recognition that, while linear MRI continues to drive the fully developed turbulence, the disc plasma is not in its undisturbed state but has a large turbulent resistivity that needs to be taken into account. Motivated by the numerical results of \citet{vaisala14} and \citet{riols17}, we assume that the turbulent resistivity, $\eta_t$, affects the behavior of the MRI in the same way as a microscopic resistivity, allowing us to apply the well-understood theory of the dissipative MRI to predict the saturation level \citep{blaes94,lesur07,pessah08}. This is distinct from models in which MRI saturation is attributed to the development of parasitic instabilities \citep{goodman94,pessah10,gogichaishvili18,hirai18}.
 
As we show in \S\ref{sec:MRI}, this interaction between MRI and the turbulent resistivity driven by the  MRI can limit the amplitude of turbulent fluctuations. 
And while the energy injection from MRI can still be dominated by the most unstable, short-wavelength modes, the turbulent resistivity and viscosity are determined by the largest available scales, e.g., the disc scale height, regardless of $\beta_z$. By setting $H$ equal to the fastest growing wavelength {\it in the presence of turbulent resistivity}, we can estimate $\eta_t$ as well as other measures of turbulence such as the viscous stress and turbulent magnetic energy density.  For standard accretion disc conditions, with $\beta_z \gg 1$ and negligible net $B_\phi$, we show that this argument leads to the empirical $\alpha \propto \beta_z^{-1/2}$ scaling. In \S\ref{sec:suprathermal}, we generalize our argument to discs containing a strong (suprathermal) toroidal magnetic field and in \S\ref{sec:elevated} we address the applicability of our model to the vertical structure of discs where magnetic support is important, particularly magnetically elevated discs.  We conclude by discussing possible tests of our model and its implications for the creation of large-scale organized disc fields through dynamo processes.  
 
\section{Self-regulation of MRI-driven turbulence}
\label{sec:MRI}
Numerical simulations show that the net-field MRI produces a saturated stress that is well-described by the relation, $\alpha \propto \beta_z^{-1/2}$, where $\alpha$ is the \citet{shakura73} viscosity parameter and $\beta_z$ is the plasma $\beta$ that corresponds to the net vertical field. We first show that this relation follows as an elementary consequence of the hypothesis of {\em MRI self-regulation}: the MRI saturates when its own turbulent resistivity would damp the linear growth of the fastest growing mode on the scale of the disc scale height. The argument relies on being able to treat the turbulent resistivity as a microscopic resistivity (so that linear results can be applied), and on weaker simulation-derived results on the structure of MRI turbulence.

\subsection{Saturation of MRI in a weakly-magnetized disc}
\label{sec:saturation}
Consider a disc with a purely vertical field characterized by an Alfv\'en speed $v_{\rm Az} = (B_{z0}^2 /4\pi\rho)^{1/2}$. In the presence of microscopic dissipation, a standard MRI analysis shows that the resistivity $\eta$ is typically more important than the viscosity and truncates MRI growth for wavenumbers such that $\eta k^2 \gtrsim \gamma_{\rm MRI}(k)$, where $\gamma_{\rm MRI}(k)$ is the growth rate of MRI in the absence of resistivity \citep{blaes94,pessah08}. The fastest-growing mode in the strong-resistivity regime becomes
\begin{equation}
    \label{kmax}
    k_{\rm max} = \sqrt{ \frac{3}{4}} \frac{v_{\rm Az}} {\eta}
\end{equation}
\citep{pessah08}.
An increased resistivity damps small scales and drives the dominant MRI mode to larger and larger wavelengths.

In well-ionized accretion discs, the microscopic resistivity is negligible but the turbulence itself generates an effective or turbulent resistivity, $\eta_t$. We adopt the results from \cite{vaisala14} and treat the effects of the turbulent transport coefficients exactly the same as if they were microscopic quantities. We then assume that the condition for saturation is that the turbulent resistivity damps the fastest growing MRI mode on the largest accessible scale, which is the disc scale height $H$. Setting $k_{\rm max} =  2\pi/H$ then allows us to estimate the saturated turbulent resistivity for net-field MRI,     
\begin{equation}
    \label{etasat}
    \eta_t = \sqrt{3\over 4} {v_{\rm Az} H\over 2\pi}.
\end{equation}
We note that this is essentially a turbulent version of the argument predicting the existence of ``dead zones'' in protoplanetary discs \citep{gammie96}. 
 
It remains to translate the predicted saturated turbulent resistivity into a saturated angular momentum transport efficiency. Defining a turbulent magnetic Prandtl number ${\rm Pr}_t \equiv \nu_t / \eta_t$, the usual definition of $\alpha$ gives,
\begin{equation}
    \alpha = \frac{3}{2} \frac{\Omega}{c_s^2} {\rm Pr}_t \eta_t.
\end{equation}
In terms of the plasma $\beta$-parameter for the net field component, $\beta_z = 8\pi \rho c_s^2 / B_{z0}^2 $, we have
\begin{equation}
    \label{alpha}
    \alpha = {3\sqrt{6}\over 8\pi} {\rm Pr}_t \beta_z^{-1/2}.
\end{equation}
The empirical relation $\alpha \propto \beta_z^{-1/2}$ is recovered provided that ${\rm Pr}_t$ is a weak function of field strength. We contend that this is a reasonable assumption. Simulations show that Pr$_t \sim 1$ on small scales within the inertial cascade \citep{guan09,lesur09,fromang09} but may be somewhat larger ($\sim 5-10$) near the outer scales considered here \citep[][and references therein]{bian21}. Equation (\ref{alpha}) thus reproduces the scaling and provides a reasonable quantitative approximation to empirical results from simulations \citep[e.g.,][]{hawley95,bai13,salvesen16,mishra20}.

The same scaling can be derived using somewhat distinct reasoning. At a fundamental level, $\eta_t$ has the scaling
\begin{equation}
    \label{etadef}
    \eta_t \sim u^2 \tau \sim {{\cal E}_t \over\rho} \tau,
\end{equation}
where $u$ is a typical velocity fluctuation, ${\cal E}_t$ is the turbulent energy density, and $\tau$ is a characteristic correlation timescale for fluctuations of a given scale $\ell$. Normally, $\tau$ is identified with the turnover time of an eddy $\sim \ell/u$, but in MRI this is true only if the eddies are not torn apart first by the background shear flow, i.e., only if $\ell/u < \Omega^{-1}$. This condition is {\it not} satisfied on scales $\sim H$ for a weakly magnetized disc, since $u \ll c_s$.
We therefore take $\tau \sim \Omega^{-1}$ such that $\eta_t \sim {\cal E}_t \rho^{-1} \Omega^{-1}$. Simulations of MRI show that the turbulent energy density scales linearly with the Maxwell stress, 
\begin{equation}
    T_{r\phi} = - \left\langle{B_r B_\phi \over 4\pi }\right\rangle,
\end{equation}
and that the Maxwell stress is the dominant contribution to the total stress. We then have that ${\cal E}_t \propto \alpha \rho c_s^2$, and equation (\ref{etasat}) again implies that $\alpha \propto \beta_z^{-1/2}$.
The simulation input invoked here is obviously equivalent to the statements regarding the turbulent Prandtl number in MRI turbulence discussed above.

\section{Saturation of suprathermal MRI}
\label{sec:suprathermal}

This argument can be generalized to include cases where MRI has a different growth rate than the standard one.  In this section, we will consider the saturation of MRI where the background magnetic energy density exceeds the gas pressure. It turns out that there is a subtlety in this case, since the presence of resistivity can actually enhance the growth of the instability.

\subsection{Importance of azimuthal modes}
\label{sec:azimuthal}

In the standard MRI configuration where $B_\phi = 0$, the instability is usually assumed to be quenched when no unstable modes can fit within $H_d$ --- this occurs for $\beta_z \lesssim 1 $.  But this assertion is invalidated by even a small organized $B_\phi$ threading the disc, since the full, local dispersion relation for MRI-like modes depends on both vertical and azimuthal wavenumbers through the dimensionless combination 
\begin{equation}
    \label{ndef}
n = {1\over \Omega} \left(k_z v_{{\rm A}z} +{m\over R} v_{{\rm A}\phi} \right), \end{equation}
where $v_{{\rm A}\phi} = (B_\phi^2 /4\pi\rho)^{1/2}$, and not on the vertical wavenumber $k_z$ by itself.  In other words, there is a degeneracy between the growth rates of purely vertical modes and modes with both an azimuthal dependence and a (different) vertical wavenumber \citep{das18,begelman22}. 

Unlike purely vertical MRI modes, which exhibit exponential growth, azimuthal and hybrid modes are transient in the sense that they possess a time-dependent radial wavenumber, $k_R$, that ultimately exceeds $k_z$, quenching growth \citep{balbus98}.  However, for our purpose this quenching effect is probably unimportant, since it occurs over a timescale, $\sim \Omega^{-1}$, which is comparable to the coherence timescale over which MRI operates on a given turbulent eddy.

In other words, in our model for saturated MRI we do not expect to see the large amplification of a given mode over even a single exponential growth time, but rather the cumulative effect  of many short-lived modes, each amplifying turbulent motions by a fractionally small amount.

Standard MRI in a Keplerian disc is quenched when $n^2 \geq 3$, but even if we have $2\pi v_{{\rm A}z} \gg \sqrt{3} \Omega H_d$ we can find unstable modes with azimuthal wavenumbers satisfying 
\begin{equation}
    \label{mrange}
-\sqrt{3} {\Omega R \over v_{{\rm A}\phi}} < m +  2\pi {R\over H_d} {v_{{\rm A}z}\over v_{{\rm A}\phi}} <  \sqrt{3} {\Omega R \over v_{{\rm A}\phi}} .
\end{equation} 
Provided that $v_{{\rm A}\phi} < 2\sqrt{3} \Omega R $, it is possible to find an integer $m$ (which may be negative) that allows instability. 

There are strong indications, from both local and global simulations, that dynamo action creates an organized, ``suprathermal'' toroidal field $B_\phi$, with $v_{{\rm A}\phi} > c_s$, even at moderately high values of $\beta_z \lesssim 10^2$, i.e., even at values of $B_z$ where the usual assumptions of MRI still hold \citep{bai13, salvesen16,mishra20}. In this case, the effects of this strong toroidal field on the growth of MRI need to be taken into account.         
 
\subsection{MRI growth in the suprathermal regime}
\label{sec:azimuthal2}

The presence of a suprathermal toroidal field affects the growth of MRI in two ways.  At ``weak'' suprathermal levels, $c_s < v_{{\rm A}\phi} < (c_s \Omega R)^{1/2}$ (modulo numerical factors that will be discussed later), the main effect is to introduce compressibility into the modes, which suppresses the MRI growth rate \citep{blaes94,pessah05}. However, for stronger fields, $(c_s \Omega R)^{1/2} < v_{{\rm A}\phi} < \Omega R$, the effects are dominated by the curvature of the field lines and radial gradients, which enhance the maximum growth rates. At still stronger field strengths, the rotation frequency $\Omega$ becomes seriously impacted by magnetic forces and radial buoyancy effects become important \citep[e.g.,][]{begelman22}; we will not consider this regime here.  All of these effects are included in a local dispersion relation presented by \cite{das18}, equation A3.  The full dispersion relation is a quartic in the frequency, but in the regime $c_s \ll v_{{\rm A}\phi} \ll \Omega R $ the fourth order term (which is critical for both the standard MRI regime and the buoyancy-dominated regime) can be neglected.  

\begin{figure}
\includegraphics[width=85mm]{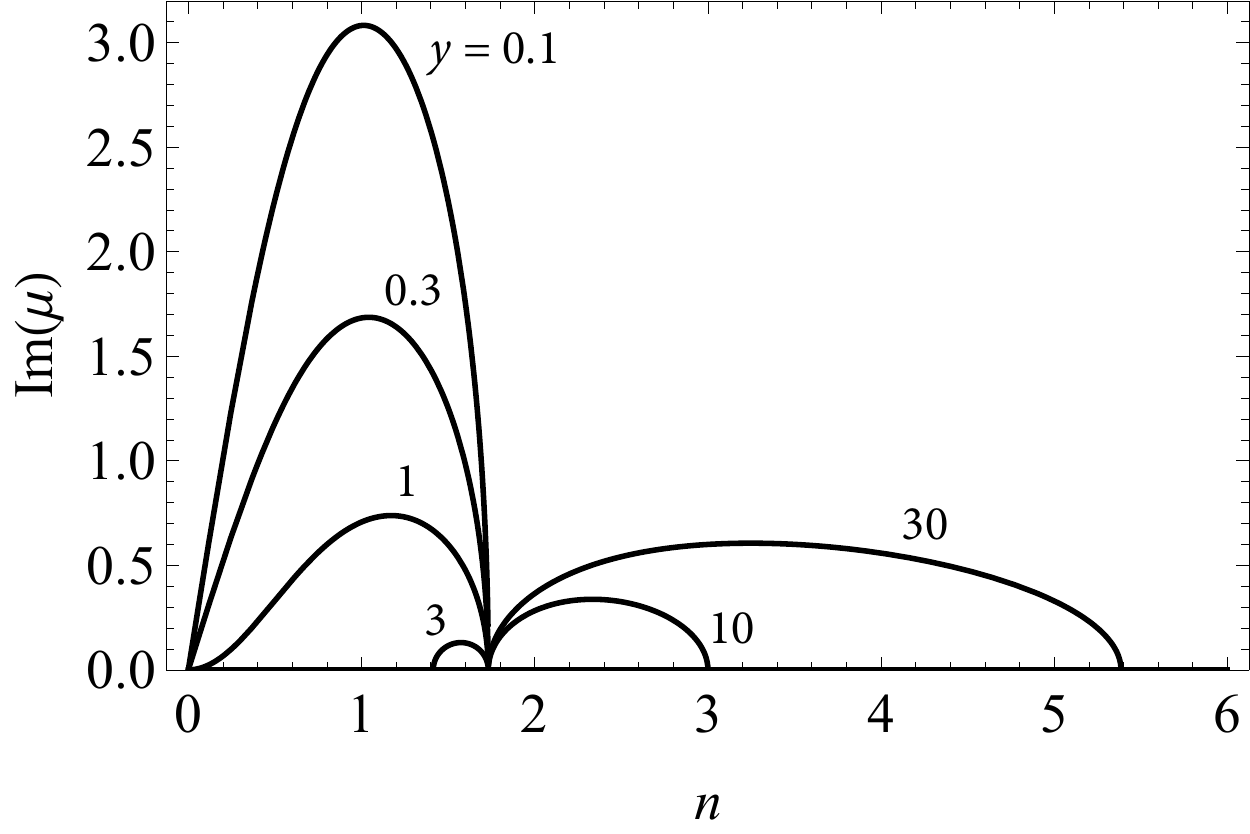}
\caption{Dimensionless growth rate Im$[\mu]$ of the most unstable mode of suprathermal MRI  (equation \ref{MRIgrowth}) in units of $v = v_{{\rm A}\phi} |1 + \hat B|$ for different values of $y$. }
\label{fig:supra}
\end{figure}

To capture the radial gradient of the toroidal field and effects of magnetic tension, we define $\hat B \equiv d\ln B_\phi / d\ln R$ and note that we still have to take into account deviations from the Keplerian rotation rate at certain places in the dispersion relation, using $(\Omega^2 - \Omega_K^2) R^2  = v_{{\rm A}\phi}^2 (1 + \hat B)$, where $\Omega_K$ is the Keplerian angular velocity.  
We can take into account a local radial wavenumber $l$ by renormalizing the MRI mode frequency $\omega$ and the generalized wavenumber $n$ (equation [\ref{ndef}]) to 
\begin{equation}
    \label{freq}
w \equiv \left(1 + {l^2\over k_z^2} \right)^{1/2} {\omega \over \Omega_K}; \ \ \ \  n \equiv  \left(1 + {l^2\over k_z^2} \right)^{1/2} {1 \over \Omega_K}\left(k_z v_{{\rm A}z} +{m\over R} v_{{\rm A}\phi} \right), 
\end{equation}
but note that the greatest instability is obtained for small $l$ so we will always assume $l\ll k_z$.  Finally, setting 
\begin{equation}
    \label{xdef}
x \equiv {c_s^2 \over v_{{\rm A}\phi}^2} ; \ \ \ \  y \equiv { v_{{\rm A}\phi}^4 (1 + \hat B)^2 \over c_s^2 v_K^2 }  , 
\end{equation}
where $v_K = R\Omega_K$ is the Keplerian speed, we obtain the suprathermal dispersion relation
\begin{equation}
    \label{supradisp}
(1 + n^2) w^2 + 4 n (x y )^{1/2}w + n^2 x (y + 3 - n^2) = 0,
\end{equation}
valid for $x < (1+y)^{-1}$, which gives the MRI growth rate
\begin{equation}
    \label{MRIgrowth}
\gamma_{\rm MRI} = {n \over (1 + n^2)}\left[ x (3-n^2)(n^2 + 1 - y) \right]^{1/2} \Omega_K 
\end{equation}
when $(3- n^2) (n^2 + 1 - y) > 0 $.  In the suprathermal limit it is convenient to express dimensionless frequencies in units of $v \equiv (x y)^{1/2} = v_{{\rm A}\phi} |1 + \hat B|/v_K$; growth rates in these units are shown as a function of $n$ in Figure \ref{fig:supra}.

Examination of equation (\ref{MRIgrowth}) shows that there are four regimes of suprathermal MRI \citep{pessah05,das18}, which we review here:
\begin{itemize}
    \item{I: $y < 1$. Instability occurs for all $0 < n^2 < 3$, as in normal MRI, but with a reduced growth rate (for $n \ll 3 $, $y\ll 1$),
 \begin{equation}
    \label{regimeI}
\gamma_{\rm MRI} \approx \sqrt{3 x} n \Omega_K \approx \sqrt{3} \left( {c_s \over v_{{\rm A}\phi}} \right) k_z v_{{\rm A}z}
\end{equation}   
The maximum growth rate for $y\ll 1$ is $\gamma_{\rm MRI} = (c_s/ v_{{\rm A}\phi}) \Omega_K$ at $n = 1$. }
    \item{II: $1 < y < 4$. MRI operates over a more restricted range of wavenumber, $y - 1 < n^2 < 3$. In the middle of this range, $n \sim O(1)$ and $\gamma_{\rm MRI} \sim v_{{\rm A}\phi}/R$ is insensitive to wavenumber. }
    \item{III: $y = 4$, corresponding to 
    \begin{equation}
    \label{regimeIII}
v_{{\rm A}\phi}^2 = {2 c_s v_K \over |1 + \hat B|} .
\end{equation}  
 For this special case there are no unstable MRI modes, i.e., $\gamma_{\rm MRI} = 0$, as discovered by \citep{pessah05} and confirmed numerically by \cite{das18}. }
    \item{IV: $y > 4$.  Instability occurs for $3 < n^2 < y - 1$, i.e., only for wavenumbers larger than standard MRI.  For $y \gg 4$, $\gamma_{\rm MRI} \approx (xy)^{1/2} \Omega_K = |1 + \hat B|v_{{\rm A}\phi}/R $, which is independent of $n$. }
\end{itemize}

We can now apply these results to estimating the saturated levels of MRI-driven turbulence in suprathermal discs.

\subsection{MRI saturation levels}
\label{sec:azimuthal3}

Except in the weakest suprathermal regime I, MRI growth rates are insensitive to the wavenumber parameter $n$.  And even for typical regime I conditions where $v_{{\rm A}\phi} \gg c_s$, we expect the vertical field to be strong enough that $2\pi v_{{\rm A}z} \gtrsim \Omega_K H_d$. We can then assume that the dominant modes contain a mixture of azimuthal and vertical wavenumbers and $\gamma_{\rm MRI}$ is close to the maximum value.  This means that, in practice, suprathermal MRI is largely independent of $n$, allowing us to combine the asymptotic growth rates for regimes I and IV to obtain   
\begin{equation}
    \label{gammasupra}
\gamma_{\rm MRI} \approx  {c_s \over v_{{\rm A}\phi}}\Omega_K \left( 1 + {v_{{\rm A}\phi}^2 \over c_s v_K} |1 + \hat B|\right). 
\end{equation}
This expression ignores the complete dropout of MRI at $y=4$, but there is no reason to believe that this condition is an attractor for disc structure and it seems reasonable to assume that most suprathermal discs will lie comfortably on either side of this limit. 

To estimate the saturation level of suprathermal MRI, we take $\eta_t k_t^2 = \gamma_{\rm MRI}$, where $k_t$ is the dominant wavenumber on which the saturation process operates.   For standard MRI, we already saw that $k_t$ is likely to correspond to the largest available scale, even though  $\gamma_{\rm MRI } \propto k$ is an increasing function of wavenumber.  This conclusion is even stronger in the limit of suprathermal $v_{{\rm A}_\phi}$, where the growth rates are less sensitive to $k$.  We therefore take $k_t = 2\pi / H_d$ for estimates within the disc core surrounding the midplane.  

Within the scale height surrounding the midplane, we then have
\begin{equation}
    \label{etasupra}
\eta_t \approx  {1 \over (2\pi)^2}\Omega_K H^2{c_s \over v_{{\rm A}\phi}} \left( 1 + {v_{{\rm A}\phi}^2 \over c_s v_K} |1 + \hat B|\right) 
\end{equation}
in the ideal MHD limit.  However, this is based on the ideal MHD dispersion relation for MRI, which suffices in the standard case with small $v_{{\rm A}\phi}$.  In the suprathermal case, we show in Appendix \ref{appendix} that a large resistivity can enhance the growth rate of MRI, as well as altering (and broadening) the range of $n$ over which instability occurs.  Evaluating the condition $\eta_t k_t^2 = \gamma_{\rm MRI}$ numerically, however, shows that the self-consistent resistivity is suppressed by a small amount, compared to the value based on the ideal dispersion relation, for $y \lesssim 3$, while being enhanced by at most a factor of $\sim 2$ for large values of $y$.  We are therefore justified in using equation (\ref{etasupra}) to make rough estimates of the saturated level of turbulence.  

As in the case of standard MRI, we can relate $\eta_t$ to the turbulent energy density through equation (\ref{etadef}), provided that we are able to estimate the coherence time $\tau$. Setting aside the effects of background shear, we are dealing with MHD turbulence in a medium with a strong guide magnetic field, in which case the coherence time for an eddy of scale $H$ is predicted to be   
\begin{equation}
    \label{tausridhar}
\tau_{\rm SG} \sim  \max \left[ 1, \left({v_{{\rm A}\phi} \over u}\right)^4 \right]  \left( H \over {v_{{\rm A}\phi} }\right)  
\end{equation}
\citep{sridhar94,goldreich95}. In our case, however, $H \gtrsim (v_{{\rm A}\phi}^2 + u^2)^{1/2} \Omega^{-1} $ due to the constraints of vertical hydrostatic equilibrium. As a result, $\tau_{\rm SG} \gtrsim \Omega^{-1}$ so the background shear dominates and we should take $\tau \sim \Omega^{-1}$.  Neglecting numerical factors, which are very uncertain, we therefore estimate the turbulent energy density to be   
\begin{equation}
    \label{Esupra}
{{\cal E}_t \over \rho} \sim  \Omega_K^2 H^2  {c_s \over v_{{\rm A}\phi} } \left( 1 + {v_{{\rm A}\phi}^2 \over c_s v_K} |1 + \hat B|\right)  \ . 
\end{equation}

\section{Magnetically elevated discs}
\label{sec:elevated}

\subsection{Inadequacy of MRI-driven turbulence}
\label{sec:inadequacy}

As noted in the Introduction, numerous simulations have found that accretion discs threaded by a net vertical flux become magnetically elevated even while the strength of the vertical magnetic field is quite small, $\beta_z \lesssim 10^2-10^3$ \citep{bai13,salvesen16,zhu18,mishra20}. Why is this? Setting aside the origin of an organized $B_\phi$ (which we discuss below, in \S~\ref{sec:dynamo}) in the first place, these simulations also suggest that the elevation cannot always be due to the organized field, which in many simulations shows a reversal near the midplane and a positive gradient within the inner (elevated) scale height that is inconsistent with vertical support against gravity.\footnote{This is the case, for example in recent global simulations by \cite{zhu18}, \cite{mishra20}, and \cite{jacqueminide21}. In contrast, simulations by \cite{huang23} show a symmetric field about the midplane that reverses frequently in time.}  Therefore, magnetic elevation --- at least near the midplane --- must be associated with sufficiently strong MHD turbulence.

One can easily see that the saturated turbulence levels we have estimated for the suprathermal regime are inadequate to this task.  Vertical equilibrium requires 
${\cal E}_t / \rho \sim  \Omega_K^2 H^2$, but equation (\ref{Esupra}) has this quantity on the right-hand side multiplied by a factor which will generally be less than one, except in the limits where $v_{{\rm A}\phi} \rightarrow c_s$ (the subthermal limit, where the disc is not elevated) and $v_{{\rm A}\phi} |1 + \hat B| \rightarrow v_K$, where the angular velocity of the disc is non-Keplerian.  The simulations indicate that magnetic elevation occurs between these two limits, where our analysis predicts that saturated MRI-driven turbulent pressure cannot support the disc vertically against gravity.  

This suggests that there must be some other source of much stronger turbulence that sets in under conditions that lead to magnetic elevation.  We propose that, given the creation of a sufficiently large $B_\phi$, the dominant instability mechanism in a disc switches from MRI to a different instability --- the tearing instability --- and that the latter is responsible for magnetic elevation.    
 
\subsection{Onset of tearing modes}
\label{sec:tearing}

Tearing instabilities affect regions of enhanced current density and could be triggered by a vertical gradient of $B_\phi$ in the presence of the turbulent resistivity produced by MRI.  Normally, tearing instabilities are associated with narrow current sheets mediated by microscopic resistivity \citep{furth63}, but here we are proposing that they operate due to turbulent resistivity across gradient length scales $\sim H$. In this regime, the Lundquist number --- the ratio of the resistive timescale to Alfv\'en transit time across the layer, $S = v_{{\rm A}\phi} H / \eta $ --- is of order a few, and we estimate the characteristic growth rate by   
\begin{equation}
    \label{gammatear}
\gamma_{\rm tear} \sim   0.1 {v_{{\rm A}\phi} \over H}   
\end{equation}
\citep{lee86}.

Tearing modes should take over from MRI as the dominant instability if $\gamma_{\rm tear} > \gamma_{\rm MRI}$.  If this is satisfied, we conjecture that the saturation of the tearing instability can be estimated the same way we have estimated the saturation of MRI, by setting $\gamma_{\rm tear} \sim \eta / H^2$, which gives\begin{equation}
    \label{etatear}
  \eta \sim 0.1 {v_{{\rm A}\phi}} H \ .
\end{equation}
Note that the Lundquist number is then $\sim 10$, which is consistent with the operation of the tearing instability \citep{lee86}.

As in the case of MRI, the background shear probably limits the correlation timescale to $\tau \sim \Omega^{-1}$, so that the turbulent energy density is 
\begin{equation}
    \label{Etear}
{{\cal E}_t \over \rho} \sim  {\eta \over \tau} \sim  0.1 v_{{\rm A}\phi}\Omega_K H \ .  
\end{equation}
If the associated turbulent pressure is required to support the disc hydrostatically, we obtain $H \sim 0.1 v_{{\rm A}\phi}/ \Omega_K$.  This does not lead to a self-consistent model if  we take the numerical coefficient literally, since it implies that the laminar pressure is much larger than the turbulent pressure and should crush the turbulent zone.  But note that we have neglected various numerical factors in relating the turbulent pressure to $\eta$, including the turbulent magnetic Prandtl number which could ameliorate the problem if it is $\sim O(10)$.  This situation is different from the turbulent pressure shortfall associated with suprathermal MRI, where the imbalance grows worse as $v_{{\rm A}\phi}/c_s$ increases.

We therefore suggest that the turbulence associated with the tearing instability, triggered by an organized suprathermal $B_\phi$, is capable in principle of magnetically elevating a disc, with turbulent and laminar magnetic pressures that are roughly in equipartition within the inner disc scale height.  We now turn our attention to factors that may set the value of $v_{{\rm A}\phi}$.

\subsection{Kinematic disc dynamo}
\label{sec:dynamo}

The organized toroidal field in a magnetically elevated disc is presumably the result of a large-scale dynamo.  Numerous numerical and analytic studies have probed the possible nature of turbulent accretion disc dynamos \citep{kato95, vishniac97, ogilvie03, gressel10, kapla11, gressel15, squire15, ebrahimi16, dhang20}, mainly in the context of mean-field models that invoke cross-correlations among various turbulent velocity components to produce the $\alpha$- and other dynamo effects\footnote{Here we are referring to the $\alpha$ pseudo-tensor of mean-field dynamo theory, and not the viscous $\alpha$-parameter of accretion disc modeling.}, including the turbulent resistivity. Most of these models have not focused on highly magnetized discs, i.e., with a net, indestructible vertical flux, which have the potentially simplifying characteristic that the $\alpha$-effect is not essential in order to create and maintain a persistent toroidal (and radial) field: with net flux and the mean velocity field (both the shear and radial flow), these fields can be maintained through purely kinematic means (although the $\alpha$-effect may very well be present and quantitatively important).  Indeed, simulations of discs with net $\beta_z$ show a lengthening of flux reversal timescales as the disc becomes more magnetized, with the cycles disappearing entirely in the most strongly magnetized cases \citep{salvesen16}.  Such cycles are often considered a hallmark of $\alpha-\omega$ dynamos, suggesting that perhaps these effects are less prominent, if not absent, in highly magnetized discs.

To estimate $B_\phi$ from purely kinematic arguments, suppose the disc develops a steady-state large-scale magnetic field, so that the mean electric field is curl-free.  Two terms that must contribute to the total electric field ${\bf E}$ are the advective term, ${\bf v} \times {\bf B}$, and the resistive term, $\eta \nabla\times {\bf B}$, where $\eta$ is the turbulent resistivity.  Other terms that may be present include those from the $\alpha$-effect, which are proportional to the mean-field ${\bf B}$, and even non-standard terms involving cross-correlations between the turbulent velocity and turbulent magnetic field components, which are proportional to components of the mean velocity field. There may also be an arbitrary potential electric field which, for example, could cause the large-scale magnetic field pattern to rotate at a different (typically, smaller) angular speed ($\Omega_p$) than the matter in the disc ($\Omega_K$).  Neglecting the $\alpha$-effect (and other related terms) but including a possible potential field in the radial component of ${\bf E}$, we can write the balance among contributions to $E_r$  as
\begin{equation}
    \label{Er}
   R\Delta \Omega B_z \sim {\eta B_\phi \over H}  \sim 0.1 {v_{{\rm A}\phi}} B_\phi \ ,  
\end{equation}
where $\Delta\Omega \equiv \Omega_K - \Omega_p$ and  we used our estimate of $\eta$ from equation (\ref{etatear}), or equivalently,
\begin{equation}
    \label{Er2}
   v_{{\rm A}\phi}^2 \sim 10 v_{{\rm A}z} v_K  \left( {\Delta\Omega \over \Omega_K } \right).  
\end{equation}
Since $E_\phi$ has to vanish in a steady-state, we also have\begin{equation}
    \label{Ephi}
   v_r B_z \sim {\eta B_r \over H} \sim 0.1 {v_{{\rm A}\phi}} B_r  \ .  
\end{equation}

The value of $B_r/B_z$ is likely determined by conditions above and below the disc, e.g., by fitting to a magnetocentrifugal wind \citep{konigl89,li95} or some other physically motivated field configuration \citep{lubow94,okuzumi14,takeuchi14}.  All of these models give a ratio of order unity, and we will adopt $B_r \sim B_z$ to obtain  
\begin{equation}
    \label{Ephi2}
   v_r  \sim {\eta \over H} \sim 0.1 {v_{{\rm A}\phi}}   \ .  
\end{equation}
For a magnetically elevated disc ($v_{{\rm A}\phi} > c_s$), this radial velocity is far larger than can be supported by internal turbulent viscosity, indicating that accretion in elevated discs is almost completely driven by large-scale magnetic torques. These torques transport the angular momentum mostly in the vertical direction, implying 
\begin{equation}
    \label{Ephi3}
   v_r  \sim {v_{{\rm A}\phi} v_{{\rm A}z} \over H \Omega_K}   \ .  
\end{equation}
A large inflow speed is no surprise, since for a thin disc and moderate Prandtl number, it is required to maintain a steady state between inward magnetic field advection and outward diffusion \citep{lubow94}.

Comparing equations (\ref{Ephi2}) and (\ref{Ephi3}), we find that the toroidal field is proportional to the vertical field, $v_{{\rm A}\phi} \sim
10 v_{{\rm A}z}$. This allows us to estimate the difference between the angular velocity of the field lines and that of the matter, using equation (\ref{Er2}): 
\begin{equation}
    \label{Er3}
 {\Delta\Omega \over \Omega_K } \sim  { v_{{\rm A}\phi}\over v_K}  \sim  {H \over R} .  
\end{equation}
Finally, we can derive a rough condition for the onset of elevated accretion by setting $v_{{\rm A}\phi} \sim 10 v_{{\rm A}z} \gtrsim c_s $ or, equivalently, $\beta_z \lesssim 100$.  

\section{Discussion and Conclusions}
\label{sec:conclusions}

We have proposed that turbulence driven by the magnetorotational instability, and possibly other instabilities in accretion discs, can saturate due to turbulent resistivity when the instability growth rate is comparable to the resistive dissipation rate, $\gamma_{\rm instab} \sim \eta k^2$, on an outer scale that typically would be comparable to the disc scale height, $k \sim H^{-1}$. Since the resistivity scales with the turbulent energy density (per unit mass), $u^2$, and a coherence time, $\tau$, we can use estimates of $\eta$ to deduce the saturated level of turbulence, provided that we can estimate $\tau$.  Normally, $\tau$ would be given by the turbulent overturn time, $(ku)^{-1}$, but in the case of an accretion disc (or other strongly sheared flow), the background shear could tear apart the eddies on a shorter timescale ($\sim \Omega^{-1}$).  We find that this is the case for MRI under most conditions, leading to larger saturated turbulence levels than would otherwise be predicted under the assumption that energy is injected on scales $\sim H$.

We first applied this saturation model to discs with a weak net vertical magnetic field $B_z$ and negligible organized toroidal field, and found that it reproduces the empirical correlation --- seen in numerous simulations but never definitively explained --- between the viscous $\alpha$-parameter and the ratio of gas pressure to vertical magnetic pressure $\beta_z$, $\alpha \propto \beta_z^{-1/2}$.  
We then applied the model to ``suprathermal'' accretion discs, in which a toroidal magnetic field ($B_\phi$) embedded in the disc exerts a pressure that exceeds the gas or radiation pressure that normally supports the disc against gravity.  We find that the presence of a significant resistivity causes some interesting modifications to the MRI growth rate (mainly discussed in Appendix \ref{appendix}), but our main conclusion is that the saturated level of MRI-driven turbulence is suppressed over much of the suprathermal regime.

Applying these results to the magnetic elevation of discs, we conclude that neither MRI-driven turbulence nor an antisymmetric, organized $B_\phi$ created by some dynamo process is able to inflate the scale height of a disc near the midplane.  However, we suggest that the tearing instability, triggered by the vertical gradient of $B_\phi$ in the presence of a modest turbulent resistivity generated initially by MRI, might lead to a level of turbulence sufficient to elevate the disc.  In a speculative spirit, we use the turbulent resistivity to estimate the saturation level of the tearing mode which, coupled with the rudimentary scaling of a kinematic dynamo, allows us to predict the onset of magnetic elevation as a function of the gas sound speed and the strength of the magnetic field threading the disc.  We find that the turbulent and organized magnetic pressures are proportional to each other, and speculate that this may provide a physically plausible closure for magnetically elevated disc dynamo models in a wider range of situations.

Continuing in a speculative spirit, we can extrapolate the arguments presented in \S\ref{sec:dynamo} to layers above the central scale height.  In order to maintain hydrostatic equilibrium through a combination of the vertical gradient of $B_\phi$ and the associated turbulence, $v_{{\rm A}\phi}$ must scale with height $\propto z$.  Since $\Delta\Omega$ is independent of $z$, equation (\ref{Er2}) then predicts that $v_{{\rm A}z} = B_z/ (4\pi \rho)^{1/2} \propto z^2$, and taking $B_z \sim$ const., we obtain the density scaling $\rho \propto z^{-4}$ for the ``wings'' of the disc.   

Our model is predicated on several assumptions that need  further study.  While there is some evidence that MRI responds to turbulent resistivity similarly to microscopic resistivity \citep{vaisala14}, these results are based on simulations of the weak-field case with zero net field; there has not been similar work on the suprathermal regime of MRI nor (to our knowledge) has it been explored in connection with the tearing instability.  And in applying our model to magnetically elevated discs, we have adopted the barest sketch of a dynamo model to estimate the large-scale field, ignoring the possible role of the $\alpha$-effect, which may be present though not absolutely necessary in discs threaded by a net vertical flux. We also did not discuss how instability is affected by  the large-scale radial field $B_r$, which must be present to satisfy Maxwell's equations in the presence of $B_z$ and which plays a key role in the dynamo. 

Given the potential for the magnetic elevation phenomenon to qualitatively affect numerous aspects of accretion disc modeling --- including thermal and gravitational stability, accretion timescales, and spectral hardness in several astrophysical contexts  \citep{begelman07,gaburov12,sadowski16,begelman17,dexter19} --- there is need for a theoretical framework to interpret the expanding library of published computational results.  We hope that the arguments sketched here will serve as a starting point for refining our understanding of how magnetic elevation works.

\section*{Acknowledgements}
We acknowledge useful discussions with Nicolas Scepi and Jason Dexter, and thank the referee for a thorough and thoughtful report that helped us to improve the paper.  This work was supported in part by NASA Astrophysics Theory Program grants NNX17AK55G and 80NSSC22K0826, and NASA TCAN award 80NSSC19K0639.

\section*{Data Availability}
No new data was generated or analyzed to support the work in this paper.




\bibliographystyle{mnras}
\bibliography{biblio} 

\begin{thebibliography}{}
\makeatletter
\relax
\def\mn@urlcharsother{\let\do\@makeother \do\$\do\&\do\#\do\^\do\_\do\%\do\~}
\def\mn@doi{\begingroup\mn@urlcharsother \@ifnextchar [ {\mn@doi@}
  {\mn@doi@[]}}
\def\mn@doi@[#1]#2{\def\@tempa{#1}\ifx\@tempa\@empty \href
  {http://dx.doi.org/#2} {doi:#2}\else \href {http://dx.doi.org/#2} {#1}\fi
  \endgroup}
\def\mn@eprint#1#2{\mn@eprint@#1:#2::\@nil}
\def\mn@eprint@arXiv#1{\href {http://arxiv.org/abs/#1} {{\tt arXiv:#1}}}
\def\mn@eprint@dblp#1{\href {http://dblp.uni-trier.de/rec/bibtex/#1.xml}
  {dblp:#1}}
\def\mn@eprint@#1:#2:#3:#4\@nil{\def\@tempa {#1}\def\@tempb {#2}\def\@tempc
  {#3}\ifx \@tempc \@empty \let \@tempc \@tempb \let \@tempb \@tempa \fi \ifx
  \@tempb \@empty \def\@tempb {arXiv}\fi \@ifundefined
  {mn@eprint@\@tempb}{\@tempb:\@tempc}{\expandafter \expandafter \csname
  mn@eprint@\@tempb\endcsname \expandafter{\@tempc}}}

\bibitem[\protect\citeauthoryear{{Bai} \& {Stone}}{{Bai} \&
  {Stone}}{2013}]{bai13}
{Bai} X.-N.,  {Stone} J.~M.,  2013, \mn@doi [\apj]
  {10.1088/0004-637X/767/1/30}, \href
  {https://ui.adsabs.harvard.edu/abs/2013ApJ...767...30B} {767, 30}

\bibitem[\protect\citeauthoryear{{Balbus} \& {Hawley}}{{Balbus} \&
  {Hawley}}{1991}]{balbus1991}
{Balbus} S.~A.,  {Hawley} J.~F.,  1991, \mn@doi [\apj] {10.1086/170270}, \href
  {https://ui.adsabs.harvard.edu/abs/1991ApJ...376..214B} {376, 214}

\bibitem[\protect\citeauthoryear{{Balbus} \& {Hawley}}{{Balbus} \&
  {Hawley}}{1998}]{balbus98}
{Balbus} S.~A.,  {Hawley} J.~F.,  1998, \mn@doi [Reviews of Modern Physics]
  {10.1103/RevModPhys.70.1}, \href
  {https://ui.adsabs.harvard.edu/abs/1998RvMP...70....1B} {70, 1}

\bibitem[\protect\citeauthoryear{{Begelman} \& {Pringle}}{{Begelman} \&
  {Pringle}}{2007}]{begelman07}
{Begelman} M.~C.,  {Pringle} J.~E.,  2007, \mn@doi [\mnras]
  {10.1111/j.1365-2966.2006.11372.x}, \href
  {https://ui.adsabs.harvard.edu/abs/2007MNRAS.375.1070B} {375, 1070}

\bibitem[\protect\citeauthoryear{{Begelman} \& {Silk}}{{Begelman} \&
  {Silk}}{2017}]{begelman17}
{Begelman} M.~C.,  {Silk} J.,  2017, \mn@doi [\mnras] {10.1093/mnras/stw2533},
  \href {https://ui.adsabs.harvard.edu/abs/2017MNRAS.464.2311B} {464, 2311}

\bibitem[\protect\citeauthoryear{{Begelman}, {Armitage}  \&
  {Reynolds}}{{Begelman} et~al.}{2015}]{begelman15}
{Begelman} M.~C.,  {Armitage} P.~J.,   {Reynolds} C.~S.,  2015, \mn@doi [\apj]
  {10.1088/0004-637X/809/2/118}, \href
  {https://ui.adsabs.harvard.edu/abs/2015ApJ...809..118B} {809, 118}

\bibitem[\protect\citeauthoryear{{Begelman}, {Scepi}  \& {Dexter}}{{Begelman}
  et~al.}{2022}]{begelman22}
{Begelman} M.~C.,  {Scepi} N.,   {Dexter} J.,  2022, \mn@doi [\mnras]
  {10.1093/mnras/stab3790}, \href
  {https://ui.adsabs.harvard.edu/abs/2022MNRAS.511.2040B} {511, 2040}

\bibitem[\protect\citeauthoryear{{Bian}, {Shang}, {Blackman}, {Collins}  \&
  {Aluie}}{{Bian} et~al.}{2021}]{bian21}
{Bian} X.,  {Shang} J.~K.,  {Blackman} E.~G.,  {Collins} G.~W.,   {Aluie} H.,
  2021, \mn@doi [\apjl] {10.3847/2041-8213/ac0fe5}, \href
  {https://ui.adsabs.harvard.edu/abs/2021ApJ...917L...3B} {917, L3}

\bibitem[\protect\citeauthoryear{{Blaes} \& {Balbus}}{{Blaes} \&
  {Balbus}}{1994}]{blaes94}
{Blaes} O.~M.,  {Balbus} S.~A.,  1994, \mn@doi [\apj] {10.1086/173634}, \href
  {https://ui.adsabs.harvard.edu/abs/1994ApJ...421..163B} {421, 163}

\bibitem[\protect\citeauthoryear{{Das}, {Begelman}  \& {Lesur}}{{Das}
  et~al.}{2018}]{das18}
{Das} U.,  {Begelman} M.~C.,   {Lesur} G.,  2018, \mn@doi [\mnras]
  {10.1093/mnras/stx2518}, \href
  {https://ui.adsabs.harvard.edu/abs/2018MNRAS.473.2791D} {473, 2791}

\bibitem[\protect\citeauthoryear{{Davis}, {Stone}  \& {Pessah}}{{Davis}
  et~al.}{2010}]{davis10}
{Davis} S.~W.,  {Stone} J.~M.,   {Pessah} M.~E.,  2010, \mn@doi [\apj]
  {10.1088/0004-637X/713/1/52}, \href
  {https://ui.adsabs.harvard.edu/abs/2010ApJ...713...52D} {713, 52}

\bibitem[\protect\citeauthoryear{{Dexter} \& {Begelman}}{{Dexter} \&
  {Begelman}}{2019}]{dexter19}
{Dexter} J.,  {Begelman} M.~C.,  2019, \mn@doi [\mnras]
  {10.1093/mnrasl/sly213}, \href
  {https://ui.adsabs.harvard.edu/abs/2019MNRAS.483L..17D} {483, L17}

\bibitem[\protect\citeauthoryear{{Dhang}, {Bendre}, {Sharma}  \&
  {Subramanian}}{{Dhang} et~al.}{2020}]{dhang20}
{Dhang} P.,  {Bendre} A.,  {Sharma} P.,   {Subramanian} K.,  2020, \mn@doi
  [\mnras] {10.1093/mnras/staa996}, \href
  {https://ui.adsabs.harvard.edu/abs/2020MNRAS.494.4854D} {494, 4854}

\bibitem[\protect\citeauthoryear{{Ebrahimi} \& {Blackman}}{{Ebrahimi} \&
  {Blackman}}{2016}]{ebrahimi16}
{Ebrahimi} F.,  {Blackman} E.~G.,  2016, \mn@doi [\mnras]
  {10.1093/mnras/stw724}, \href
  {https://ui.adsabs.harvard.edu/abs/2016MNRAS.459.1422E} {459, 1422}

\bibitem[\protect\citeauthoryear{{Fromang} \& {Stone}}{{Fromang} \&
  {Stone}}{2009}]{fromang09}
{Fromang} S.,  {Stone} J.~M.,  2009, \mn@doi [\aap]
  {10.1051/0004-6361/200912752}, \href
  {https://ui.adsabs.harvard.edu/abs/2009A&A...507...19F} {507, 19}

\bibitem[\protect\citeauthoryear{{Furth}, {Killeen}  \& {Rosenbluth}}{{Furth}
  et~al.}{1963}]{furth63}
{Furth} H.~P.,  {Killeen} J.,   {Rosenbluth} M.~N.,  1963, \mn@doi [Physics of
  Fluids] {10.1063/1.1706761}, \href
  {https://ui.adsabs.harvard.edu/abs/1963PhFl....6..459F} {6, 459}

\bibitem[\protect\citeauthoryear{{Gaburov}, {Johansen}  \& {Levin}}{{Gaburov}
  et~al.}{2012}]{gaburov12}
{Gaburov} E.,  {Johansen} A.,   {Levin} Y.,  2012, \mn@doi [\apj]
  {10.1088/0004-637X/758/2/103}, \href
  {https://ui.adsabs.harvard.edu/abs/2012ApJ...758..103G} {758, 103}

\bibitem[\protect\citeauthoryear{{Gammie}}{{Gammie}}{1996}]{gammie96}
{Gammie} C.~F.,  1996, \mn@doi [\apj] {10.1086/176735}, \href
  {https://ui.adsabs.harvard.edu/abs/1996ApJ...457..355G} {457, 355}

\bibitem[\protect\citeauthoryear{{Gogichaishvili}, {Mamatsashvili}, {Horton}
  \& {Chagelishvili}}{{Gogichaishvili} et~al.}{2018}]{gogichaishvili18}
{Gogichaishvili} D.,  {Mamatsashvili} G.,  {Horton} W.,   {Chagelishvili} G.,
  2018, \mn@doi [\apj] {10.3847/1538-4357/aadbad}, \href
  {https://ui.adsabs.harvard.edu/abs/2018ApJ...866..134G} {866, 134}

\bibitem[\protect\citeauthoryear{{Goldreich} \& {Sridhar}}{{Goldreich} \&
  {Sridhar}}{1995}]{goldreich95}
{Goldreich} P.,  {Sridhar} S.,  1995, \mn@doi [\apj] {10.1086/175121}, \href
  {https://ui.adsabs.harvard.edu/abs/1995ApJ...438..763G} {438, 763}

\bibitem[\protect\citeauthoryear{{Goodman} \& {Xu}}{{Goodman} \&
  {Xu}}{1994}]{goodman94}
{Goodman} J.,  {Xu} G.,  1994, \mn@doi [\apj] {10.1086/174562}, \href
  {https://ui.adsabs.harvard.edu/abs/1994ApJ...432..213G} {432, 213}

\bibitem[\protect\citeauthoryear{{Gressel}}{{Gressel}}{2010}]{gressel10}
{Gressel} O.,  2010, \mn@doi [\mnras] {10.1111/j.1365-2966.2010.16440.x}, \href
  {https://ui.adsabs.harvard.edu/abs/2010MNRAS.405...41G} {405, 41}

\bibitem[\protect\citeauthoryear{{Gressel} \& {Pessah}}{{Gressel} \&
  {Pessah}}{2015}]{gressel15}
{Gressel} O.,  {Pessah} M.~E.,  2015, \mn@doi [\apj]
  {10.1088/0004-637X/810/1/59}, \href
  {https://ui.adsabs.harvard.edu/abs/2015ApJ...810...59G} {810, 59}

\bibitem[\protect\citeauthoryear{{Guan} \& {Gammie}}{{Guan} \&
  {Gammie}}{2009}]{guan09}
{Guan} X.,  {Gammie} C.~F.,  2009, \mn@doi [\apj]
  {10.1088/0004-637X/697/2/1901}, \href
  {https://ui.adsabs.harvard.edu/abs/2009ApJ...697.1901G} {697, 1901}

\bibitem[\protect\citeauthoryear{{Hawley}, {Gammie}  \& {Balbus}}{{Hawley}
  et~al.}{1995}]{hawley95}
{Hawley} J.~F.,  {Gammie} C.~F.,   {Balbus} S.~A.,  1995, \apj, 440, 742

\bibitem[\protect\citeauthoryear{{Hirai}, {Katoh}, {Terada}  \&
  {Kawai}}{{Hirai} et~al.}{2018}]{hirai18}
{Hirai} K.,  {Katoh} Y.,  {Terada} N.,   {Kawai} S.,  2018, \mn@doi [\apj]
  {10.3847/1538-4357/aaa5b2}, \href
  {https://ui.adsabs.harvard.edu/abs/2018ApJ...853..174H} {853, 174}

\bibitem[\protect\citeauthoryear{{Huang}, {Jiang}, {Feng}, {Davis}, {Stone}  \&
  {Middleton}}{{Huang} et~al.}{2023}]{huang23}
{Huang} J.,  {Jiang} Y.-F.,  {Feng} H.,  {Davis} S.~W.,  {Stone} J.~M.,
  {Middleton} M.~J.,  2023, \mn@doi [\apj] {10.3847/1538-4357/acb6fc}, \href
  {https://ui.adsabs.harvard.edu/abs/2023ApJ...945...57H} {945, 57}

\bibitem[\protect\citeauthoryear{{Jacquemin-Ide}, {Lesur}  \&
  {Ferreira}}{{Jacquemin-Ide} et~al.}{2021}]{jacqueminide21}
{Jacquemin-Ide} J.,  {Lesur} G.,   {Ferreira} J.,  2021, \mn@doi [\aap]
  {10.1051/0004-6361/202039322}, \href
  {https://ui.adsabs.harvard.edu/abs/2021A&A...647A.192J} {647, A192}

\bibitem[\protect\citeauthoryear{{Jiang} \& {Blaes}}{{Jiang} \&
  {Blaes}}{2020}]{jiang20}
{Jiang} Y.-F.,  {Blaes} O.,  2020, \mn@doi [\apj] {10.3847/1538-4357/aba4b7},
  \href {https://ui.adsabs.harvard.edu/abs/2020ApJ...900...25J} {900, 25}

\bibitem[\protect\citeauthoryear{{K{\"a}pyl{\"a}} \& {Korpi}}{{K{\"a}pyl{\"a}}
  \& {Korpi}}{2011}]{kapla11}
{K{\"a}pyl{\"a}} P.~J.,  {Korpi} M.~J.,  2011, \mn@doi [\mnras]
  {10.1111/j.1365-2966.2010.18184.x}, \href
  {https://ui.adsabs.harvard.edu/abs/2011MNRAS.413..901K} {413, 901}

\bibitem[\protect\citeauthoryear{{Kato} \& {Yoshizawa}}{{Kato} \&
  {Yoshizawa}}{1995}]{kato95}
{Kato} S.,  {Yoshizawa} A.,  1995, \pasj, \href
  {https://ui.adsabs.harvard.edu/abs/1995PASJ...47..629K} {47, 629}

\bibitem[\protect\citeauthoryear{{K\"onigl}}{{K\"onigl}}{1989}]{konigl89}
{K\"onigl} A.,  1989, \mn@doi [\apj] {10.1086/167585}, \href
  {https://ui.adsabs.harvard.edu/abs/1989ApJ...342..208K} {342, 208}

\bibitem[\protect\citeauthoryear{{Lan{\v{c}}ov{\'a}}
  et~al.,}{{Lan{\v{c}}ov{\'a}} et~al.}{2019}]{lancova19}
{Lan{\v{c}}ov{\'a}} D.,  et~al., 2019, \mn@doi [\apjl]
  {10.3847/2041-8213/ab48f5}, \href
  {https://ui.adsabs.harvard.edu/abs/2019ApJ...884L..37L} {884, L37}

\bibitem[\protect\citeauthoryear{{Latter}, {Fromang}  \& {Faure}}{{Latter}
  et~al.}{2015}]{latter15}
{Latter} H.~N.,  {Fromang} S.,   {Faure} J.,  2015, \mn@doi [\mnras]
  {10.1093/mnras/stv1890}, \href
  {https://ui.adsabs.harvard.edu/abs/2015MNRAS.453.3257L} {453, 3257}

\bibitem[\protect\citeauthoryear{{Lee} \& {Fu}}{{Lee} \& {Fu}}{1986}]{lee86}
{Lee} L.~C.,  {Fu} Z.~F.,  1986, \mn@doi [\jgr] {10.1029/JA091iA03p03311},
  \href {https://ui.adsabs.harvard.edu/abs/1986JGR....91.3311L} {91, 3311}

\bibitem[\protect\citeauthoryear{{Lesur} \& {Longaretti}}{{Lesur} \&
  {Longaretti}}{2007}]{lesur07}
{Lesur} G.,  {Longaretti} P.~Y.,  2007, \mn@doi [\mnras]
  {10.1111/j.1365-2966.2007.11888.x}, \href
  {https://ui.adsabs.harvard.edu/abs/2007MNRAS.378.1471L} {378, 1471}

\bibitem[\protect\citeauthoryear{{Lesur} \& {Longaretti}}{{Lesur} \&
  {Longaretti}}{2009}]{lesur09}
{Lesur} G.,  {Longaretti} P.~Y.,  2009, \mn@doi [\aap]
  {10.1051/0004-6361/200912272}, \href
  {https://ui.adsabs.harvard.edu/abs/2009A&A...504..309L} {504, 309}

\bibitem[\protect\citeauthoryear{{Li}}{{Li}}{1995}]{li95}
{Li} Z.-Y.,  1995, \mn@doi [\apj] {10.1086/175657}, \href
  {https://ui.adsabs.harvard.edu/abs/1995ApJ...444..848L} {444, 848}

\bibitem[\protect\citeauthoryear{Lubow, Papaloizou  \& Pringle}{Lubow
  et~al.}{1994}]{lubow94}
Lubow S.~H.,  Papaloizou J. C.~B.,   Pringle J.~E.,  1994, \mn@doi [\mnras]
  {10.1093/mnras/267.2.235}, 267, 235

\bibitem[\protect\citeauthoryear{{Mishra}, {Begelman}, {Armitage}  \&
  {Simon}}{{Mishra} et~al.}{2020}]{mishra20}
{Mishra} B.,  {Begelman} M.~C.,  {Armitage} P.~J.,   {Simon} J.~B.,  2020,
  \mn@doi [\mnras] {10.1093/mnras/stz3572}, \href
  {https://ui.adsabs.harvard.edu/abs/2020MNRAS.492.1855M} {492, 1855}

\bibitem[\protect\citeauthoryear{{Ogilvie}}{{Ogilvie}}{2003}]{ogilvie03}
{Ogilvie} G.~I.,  2003, \mn@doi [\mnras] {10.1046/j.1365-8711.2003.06359.x},
  \href {https://ui.adsabs.harvard.edu/abs/2003MNRAS.340..969O} {340, 969}

\bibitem[\protect\citeauthoryear{{Okuzumi}, {Takeuchi}  \& {Muto}}{{Okuzumi}
  et~al.}{2014}]{okuzumi14}
{Okuzumi} S.,  {Takeuchi} T.,   {Muto} T.,  2014, \mn@doi [\apj]
  {10.1088/0004-637X/785/2/127}, \href
  {https://ui.adsabs.harvard.edu/abs/2014ApJ...785..127O} {785, 127}

\bibitem[\protect\citeauthoryear{{Pessah}}{{Pessah}}{2010}]{pessah10}
{Pessah} M.~E.,  2010, \mn@doi [\apj] {10.1088/0004-637X/716/2/1012}, \href
  {https://ui.adsabs.harvard.edu/abs/2010ApJ...716.1012P} {716, 1012}

\bibitem[\protect\citeauthoryear{{Pessah} \& {Chan}}{{Pessah} \&
  {Chan}}{2008}]{pessah08}
{Pessah} M.~E.,  {Chan} C.-k.,  2008, \mn@doi [\apj] {10.1086/589915}, \href
  {https://ui.adsabs.harvard.edu/abs/2008ApJ...684..498P} {684, 498}

\bibitem[\protect\citeauthoryear{{Pessah} \& {Psaltis}}{{Pessah} \&
  {Psaltis}}{2005}]{pessah05}
{Pessah} M.~E.,  {Psaltis} D.,  2005, \mn@doi [\apj] {10.1086/430940}, \href
  {https://ui.adsabs.harvard.edu/abs/2005ApJ...628..879P} {628, 879}

\bibitem[\protect\citeauthoryear{{Riols}, {Rincon}, {Cossu}, {Lesur}, {Ogilvie}
   \& {Longaretti}}{{Riols} et~al.}{2017}]{riols17}
{Riols} A.,  {Rincon} F.,  {Cossu} C.,  {Lesur} G.,  {Ogilvie} G.~I.,
  {Longaretti} P.~Y.,  2017, \mn@doi [\aap] {10.1051/0004-6361/201629285},
  \href {https://ui.adsabs.harvard.edu/abs/2017A&A...598A..87R} {598, A87}

\bibitem[\protect\citeauthoryear{{Ryan}, {Gammie}, {Fromang}  \&
  {Kestener}}{{Ryan} et~al.}{2017}]{ryan17}
{Ryan} B.~R.,  {Gammie} C.~F.,  {Fromang} S.,   {Kestener} P.,  2017, \mn@doi
  [\apj] {10.3847/1538-4357/aa6a52}, \href
  {https://ui.adsabs.harvard.edu/abs/2017ApJ...840....6R} {840, 6}

\bibitem[\protect\citeauthoryear{{Salvesen}, {Simon}, {Armitage}  \&
  {Begelman}}{{Salvesen} et~al.}{2016}]{salvesen16}
{Salvesen} G.,  {Simon} J.~B.,  {Armitage} P.~J.,   {Begelman} M.~C.,  2016,
  \mn@doi [\mnras] {10.1093/mnras/stw029}, \href
  {https://ui.adsabs.harvard.edu/abs/2016MNRAS.457..857S} {457, 857}

\bibitem[\protect\citeauthoryear{{Scepi}, {Lesur}, {Dubus}  \& {Flock}}{{Scepi}
  et~al.}{2018}]{scepi18}
{Scepi} N.,  {Lesur} G.,  {Dubus} G.,   {Flock} M.,  2018, \mn@doi [\aap]
  {10.1051/0004-6361/201833921}, \href
  {https://ui.adsabs.harvard.edu/abs/2018A&A...620A..49S} {620, A49}

\bibitem[\protect\citeauthoryear{{Shakura} \& {Sunyaev}}{{Shakura} \&
  {Sunyaev}}{1973}]{shakura73}
{Shakura} N.~I.,  {Sunyaev} R.~A.,  1973, \aap, \href
  {https://ui.adsabs.harvard.edu/abs/1973A&A....24..337S} {24, 337}

\bibitem[\protect\citeauthoryear{{Simon}, {Beckwith}  \& {Armitage}}{{Simon}
  et~al.}{2012}]{simon12}
{Simon} J.~B.,  {Beckwith} K.,   {Armitage} P.~J.,  2012, \mn@doi [\mnras]
  {10.1111/j.1365-2966.2012.20835.x}, \href
  {https://ui.adsabs.harvard.edu/abs/2012MNRAS.422.2685S} {422, 2685}

\bibitem[\protect\citeauthoryear{{S{\k{a}}dowski}}{{S{\k{a}}dowski}}{2016}]{sadowski16}
{S{\k{a}}dowski} A.,  2016, \mn@doi [\mnras] {10.1093/mnras/stw913}, \href
  {https://ui.adsabs.harvard.edu/abs/2016MNRAS.459.4397S} {459, 4397}

\bibitem[\protect\citeauthoryear{{Squire} \& {Bhattacharjee}}{{Squire} \&
  {Bhattacharjee}}{2015}]{squire15}
{Squire} J.,  {Bhattacharjee} A.,  2015, \mn@doi [\prl]
  {10.1103/PhysRevLett.115.175003}, \href
  {https://ui.adsabs.harvard.edu/abs/2015PhRvL.115q5003S} {115, 175003}

\bibitem[\protect\citeauthoryear{{Sridhar} \& {Goldreich}}{{Sridhar} \&
  {Goldreich}}{1994}]{sridhar94}
{Sridhar} S.,  {Goldreich} P.,  1994, \mn@doi [\apj] {10.1086/174600}, \href
  {https://ui.adsabs.harvard.edu/abs/1994ApJ...432..612S} {432, 612}

\bibitem[\protect\citeauthoryear{{Takeuchi} \& {Okuzumi}}{{Takeuchi} \&
  {Okuzumi}}{2014}]{takeuchi14}
{Takeuchi} T.,  {Okuzumi} S.,  2014, \mn@doi [\apj]
  {10.1088/0004-637X/797/2/132}, \href
  {https://ui.adsabs.harvard.edu/abs/2014ApJ...797..132T} {797, 132}

\bibitem[\protect\citeauthoryear{{V{\"a}is{\"a}l{\"a}}, {Brandenburg}, {Mitra},
  {K{\"a}pyl{\"a}}  \& {Mantere}}{{V{\"a}is{\"a}l{\"a}}
  et~al.}{2014}]{vaisala14}
{V{\"a}is{\"a}l{\"a}} M.~S.,  {Brandenburg} A.,  {Mitra} D.,  {K{\"a}pyl{\"a}}
  P.~J.,   {Mantere} M.~J.,  2014, \mn@doi [\aap]
  {10.1051/0004-6361/201322837}, \href
  {https://ui.adsabs.harvard.edu/abs/2014A&A...567A.139V} {567, A139}

\bibitem[\protect\citeauthoryear{{Vishniac} \& {Brandenburg}}{{Vishniac} \&
  {Brandenburg}}{1997}]{vishniac97}
{Vishniac} E.~T.,  {Brandenburg} A.,  1997, \mn@doi [\apj] {10.1086/303504},
  \href {https://ui.adsabs.harvard.edu/abs/1997ApJ...475..263V} {475, 263}

\bibitem[\protect\citeauthoryear{{Zhu} \& {Stone}}{{Zhu} \&
  {Stone}}{2018}]{zhu18}
{Zhu} Z.,  {Stone} J.~M.,  2018, \mn@doi [\apj] {10.3847/1538-4357/aaafc9},
  \href {https://ui.adsabs.harvard.edu/abs/2018ApJ...857...34Z} {857, 34}

\makeatother
\end{thebibliography}

\onecolumn
\appendix
\section{Resistive MRI in the suprathermal limit}
\label{appendix}

It is straightforward, albeit algebraically tedious, to modify the analysis of \cite{das18} (hereafter DBL18) to include resistivity.  In the basic set of perturbed equations (DBL18 equations 9-15), the only changes are to replace the frequency $\omega$ by $\omega + i\eta k^2$ in the three components of the magnetic induction equation (DBL18 equations 13-15).  In the momentum and continuity equations the frequency is unchanged.  Using the normalized variables defined in DBL18 Table 1, we define a resistive frequency, 
\begin{equation}
\label{chidef}
\chi \equiv \omega - m\Omega + i\eta k^2 = \mu + i \eta k^2, \end{equation} 
where $\mu$ is the scaled frequency used in DBL18. We then repeat the derivation in DBL18, keeping track of the distinction between $\chi$ and $\mu$. To avoid confusion with our parameter $y$ as defined in  equation (\ref{xdef}), we replace the DBL18 quantity $2 + y$ by $h$, but we note that $n$ in equations (\ref{DBL38}) and (\ref{DBL39}) below is normalized as in DBL18, not according to our equation (\ref{ndef}). In the penultimate step in the derivation of the dispersion relation, the resistive versions of DBL equations 38 and 39 take the form 
\begin{equation}
\label{DBL38}
\biggl\{ \frac{1}{2} (x \mu\chi - n^2) \biggl( 1 + \frac{l^2}{k_z^2}  \biggr)   + 
 \frac{h}{2} \biggl({\chi\over\mu}(1 + \hat{B}) \left[ 1 + x \left({\mu\over \chi} - 1\right) \right]
+  \frac{2-q}{n} x \tilde{\Omega} \chi \biggr)    \biggr\}  u_r   
- \biggl[ \tilde{\Omega}   + \frac{\mu h}{2 n} \biggr] x \chi (i u_\phi) = 0 
\end{equation}
and
\begin{equation}
\label{DBL39}
- i u_\phi =  \Biggl\{ \frac{  n h + \left(1+{\chi\over\mu} x\right)(2-q) \tilde{\Omega} \mu   
+ \frac{n^2}{\chi}q \tilde{\Omega} 
+ (1 + \hat B)n\left[{\chi\over\mu} + {\mu\over\chi} + x \left(1 - {\chi\over\mu}\right) - 2 \right]  }{n^2 - \left(1+{\chi\over\mu}x\right) \mu^2}   \Biggr\}  u_r ~,
\end{equation}
which reduce to their DBL18 counterparts when $\chi = \mu$.  These equations can be combined to yield the resistive version of the dispersion relation, DBL18 equation 40.  (Note that in order to explore resistive effects in the strongly magnetized limit, it is not necessary to use the more complete dispersion relation, DBL18 equation A3, which includes buoyancy effects due to gas pressure.)  

To simplify further, we define
\begin{equation}
    \label{fdef}
    f\equiv 1+{\chi\over\mu}x; \ \ \ \ g \equiv {\chi\over\mu} + {\mu\over\chi} + x \left(1 - {\chi\over\mu}\right) - 2,
\end{equation}
take the limit $l^2 / k_z^2 \rightarrow 0$ and set $q = 3/2$. We specialize to the suprathermal limit ($x\ll 1$) by setting $h = (1+ \hat B)/x$ and we define a new parameter,
\begin{equation}
    \label{vdef}
    v \equiv v_{{\rm A}\phi} (1 + \hat B). 
\end{equation}
To convert to Keplerian units (in which velocities are normalized to $v_K$, etc.), we multiply the dispersion relation in DBL18 units by $c_s^2 v_{{\rm A}\phi}^2$.  The result is
\begin{align}
\label{DBL40}
- f \mu^3 \chi + f \left\{ n^2 - {\chi\over\mu}{v^2 \over x} \left[ 1 + x \left({\mu\over \chi} - 1\right) \right]\right\} \mu^2 + \left( {v^2 \over x} + v^2 g + f \Omega^2 + x n^2 \right) \mu \chi & + n v \Omega \left[{3\over 2}\mu + \left( {5\over 2} + 2gx\right)\chi \right] 
\nonumber \\  & + n^2 \left\{ (3 \Omega^2 - n^2)x + {\chi\over\mu} v^2 \left[  1 + x \left( {\mu\over\chi} -1\right)\right] \right\} = 0 \ ,
\end{align}
where $n$ is now given by equation (\ref{ndef}).

Equation (\ref{DBL40}) is valid in the subthermal, Keplerian limit with $\Omega = 1$ and $x \gg 1$; in this case we ignore terms containing $v$ and set $f = \chi x/\mu$, yielding 
\begin{equation}
    \label{PessahChan}
    \mu^2\chi^2 - (\chi^2 + 2 n^2 \mu\chi) + n^2(n^2 - 3) =0,  
\end{equation}
in agreement with \cite{pessah08}, equation 25.  Results from this dispersion relation form the basis for our analysis in \S\ref{sec:saturation}.  To obtain the suprathermal limit, we neglect the quartic term ($\propto \mu^3\chi$) and take $x \ll 1$ and $f = 1$.  We can also assume that $g \sim O(1)$ so that $g x \ll 1$ and we can ignore all terms that depend on $g$ in equation (\ref{DBL40}).  We also assume $x n^2 \ll 1$, allowing us to  neglect the $xn^2 \mu\chi$ term.  The approximate suprathermal dispersion relation is then
\begin{equation}
    \label{supraresist}
    \mu\chi + n^2\mu^2 + nv \left({3\over 2} \mu + {5\over 2} \chi\right) + n^2\left[(3 - n^2) x + {\chi\over\mu} v^2\right] =0 \ ,
\end{equation}
which reduces to the suprathermal case without resistivity, equation (\ref{supradisp}), when $\chi = \mu$.  To unpack the implications of this dispersion relation we use equation (\ref{chidef}) to substitute $\chi = \mu + ij$, where $j \equiv \eta k^2$ is real and positive.  Note that the resistive term depends on $k^2 = k_z^2 + m^2$, while $n \propto {\bf k} \cdot {\bf v_{\rm A}} = k_z v_{{\rm A}z} + m v_{{\rm A}\phi} $, so there is no straightforward relationship between $j$ and $n$; therefore we treat them as independent variables. The resulting dispersion relation is
\begin{equation}
    \label{supraresist2}
    (1 + n^2)\mu^2 + (4 n + ij) \mu + n^2\left[(3 - n^2)x + v^2 \left( 1 + i {j\over \mu} \right) + {5\over 2} i {v j\over n } \right] =0 \ .
\end{equation}
Finally, it is convenient to use $v$ rather than $v_K$ to normalize $\mu$ and $j$. Dividing equation (\ref{supraresist2}) by $v^2$ and adopting the new scalings, we obtain
\begin{equation}
    \label{supraresist3}
    (1 + n^2)\mu^3 + (4 n + ij) \mu^2 + n^2\left[{(3 - n^2)\over y} + 1  + {5\over 2}i {j\over n} \right]\mu + i j n^2  =0 \ ,
\end{equation}
where $y \equiv v^2 / x$ as defined in equation (\ref{xdef}).

\subsection{Strongly resistive limit}
\label{sec:strongres}

\begin{figure}
\includegraphics[width=85mm]{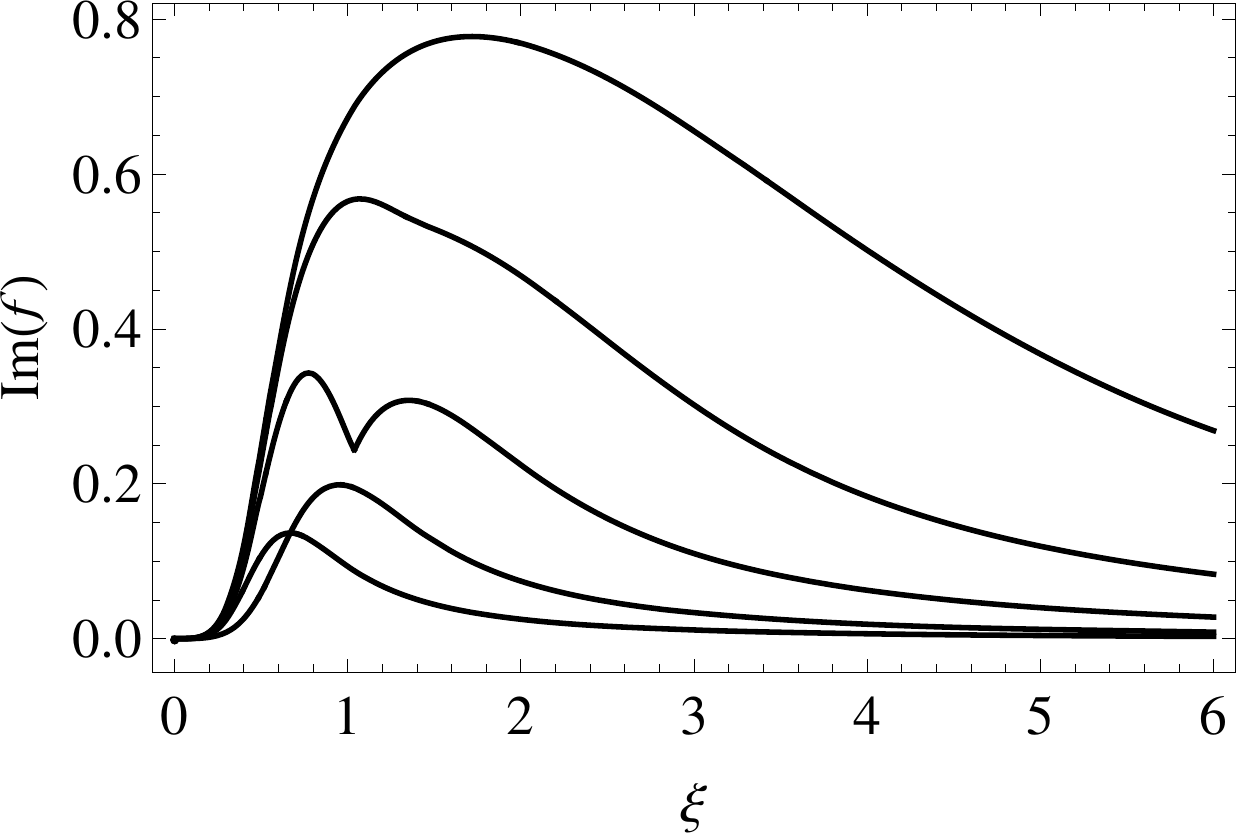}
\caption{Scaled growth rate Im$[f(\xi)]$ of the most unstable mode of suprathermal MRI  (equation \ref{fxi}) in the strongly resistive limit ($j \gg 1$), for magnetization parameters (top to bottom curve) $y = 10,3,1,0.3,0.1$. }
\label{fig:fxi}
\end{figure}

In the asymptotic limit $j \gg 1$, $n \gg1$, the scaled frequency takes on a self-similar form
\begin{equation}
    \label{fxidef}
 \mu = j^{1/3} f(\xi) \ \ \ {\rm where} \ \ \ \xi \equiv {n\over j^{1/3}} \ .
\end{equation}
In this limit, the dispersion relation reads
\begin{equation}
    \label{fxi}
    \xi^2 f^3 + i f^2 + \left( {5\over 2}i \xi - {\xi^4 \over y}  \right)f + i \xi^2   =0 \ ;
\end{equation}
solutions of equation (\ref{fxi}) for different $y$ are shown in Figure \ref{fig:fxi}. These results show that a large resistivity actually enhances the growth rate of MRI in the suprathermal regime. However, as we show in \S\ref{sec:azimuthal3}, this effect is not large enough to enhance the self-consistent saturation level of the MRI-driven turbulence.

\bsp	
\label{lastpage}
\end{document}